\documentclass[12pt]{article}
\usepackage{a4wide,graphicx,amsmath,amssymb,thophys}
\newcommand{\bs}[1]{\mbox{\boldmath ${#1}$}}
\newlength{\orbarwd}\newlength{\orbarht}\newsavebox{\orbararg}%
\newcommand{\orbar}[1]{\savebox{\orbararg}{\ensuremath{#1}}%
  \settowidth{\orbarwd}{\usebox{\orbararg}}%
  \settoheight{\orbarht}{\usebox{\orbararg}}%
  \raisebox{1.1\orbarht}{\makebox[0pt][l]{%
    \resizebox{\orbarwd}{0.5ex}{\boldmath\ensuremath{(-)}}}}%
  \usebox{\orbararg}}
\begin{document}
\begin{titlepage}

\begin{flushright}
TTP00-22\\
hep-ph/0010217\\
October 2000\\
\end{flushright}
\vskip 4.0ex
\begin{center}
\boldmath
{\Large\bf Kinematic Effects in \\[0.5ex] Radiative Quarkonia Decays}
\unboldmath
\vskip 6.0ex
{\sc Stefan Wolf}
\vskip 2.0ex
{\em Institut f\"ur Theoretische Teilchenphysik, Universit\"at Karlsruhe,\\
D-76128 Karlsruhe, Germany}
\end{center}

\begin{abstract}
\noindent
Non-relativistic QCD (NRQCD) predicts colour octet contributions to be
significant not only in many production processes of heavy quarkonia but also
in their radiative decays. We investigate the photon energy distributions in
these processes in the endpoint region. There the velocity expansion of NRQCD
breaks down which requires a resummation of an infinite class of colour octet
operators to so-called shape functions. We model these non-perturbative
functions by the emission of a soft gluon cluster in the initial state. We
found that the spectrum in the endpoint region is poorly understood if the
values for the colour octet matrix elements are taken as large as indicated
from NRQCD scaling rules. Therefore the endpoint region should not be taken
into account for a fit of the strong coupling constant at the scale of the
heavy quark mass.
\\[2ex]
\end{abstract}

\vfill

\end{titlepage}

\section{Introduction}

Since the discovery of the $J/\psi$ \cite{Aubert:1974jsAugustin:1974xw} and the
$\Upsilon$ \cite{Herb:1977ek} heavy quarkonia decays are one of the most
interesting laboratories for investigations within the framework of
perturbative QCD. In particular these bound states of a heavy quark $Q$ and its
antiparticle $\overline{Q}$ have been examined to extract the value of the
strong coupling constant $\alpha_s$ at the scale of the heavy quark mass $m_Q$.

Early theoretical analyses starting from the calculation of the total rates in
leptonic and inclusive hadronic decays \cite{Appelquist:1975zd} were done in
the colour singlet model (CSM). It {\it assumes} the quark-antiquark pair being
in the same quantum state $n = {}^{2S+1}\!L_J^{(C)}$ on the partonic level as
the corresponding quarkonium on the hadronic level. In particular the
$Q\overline{Q}$ pair has to be in a colour singlet state ($C = 1$) when it
annihilates. As a consequence of this requirement the underlying partonic
process in the radiative decay $H \to \gamma X$ of a quarkonium $H$ in the
ground state ${}^3\!S_1$ is the annihilation of the heavy $Q\overline{Q}$ pair
into a photon and at least two gluons. This process was calculated first in
\cite{Brodsky:1978du}.

Great theoretical progress in the understanding of bound states of heavy
$Q\overline{Q}$ systems has been achieved by NRQCD (Non-Relativistic Quantum
Chromo-Dynamics) \cite{Bodwin:1995jh}. In this {\it theory} quarkonia decays
are factorized into two step processes: the short-distance annihilation of a
$Q\overline{Q}$ pair with fixed total spin $S$, orbital angular momentum $L$,
and total angular momentum $J$ and its preceding long-distance transition into
this state. While the partonic subprocess can be calculated perturbatively to
definite order in $\alpha_s$ the non-perturbative subprocess $H \to
Q\overline{Q}[n] +$ {\it soft degrees of freedom} is parameterized by NRQCD
matrix elements. They are the NRQCD counterparts of the wave function at the
origin in the colour singlet model and give the probability for finding the
quark-antiquark pair in the quantum state $n$ at the moment of annihilation. In
principle the values of these parameters are unknown and must be fitted to
experimental data \cite{Braaten:2000cm} or computed on the lattice
\cite{Bodwin:1996tg}. Nevertheless NRQCD provides scaling rules which
e.g.~predict colour octet matrix elements being suppressed by powers of the
non-relativistic velocity $v$ with respect to the leading order colour singlet
matrix element. This typical velocity of the heavy (anti)quark inside the
quarkonium simultaneously serves as expansion parameter of the effective field
theory. As a result NRQCD describes a decay rate by an infinite sum over matrix
elements of four fermion operators with Wilson coefficients which on their part
are expansions in $\alpha_s$.

\begin{figure}[t]
\begin{center}
 \leavevmode
 \includegraphics[]{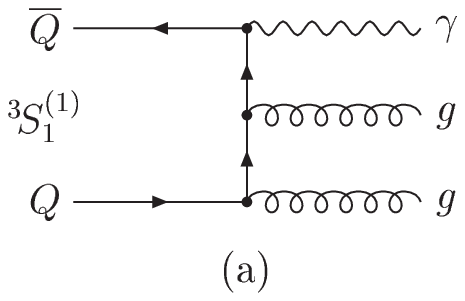}
 \qquad \qquad
 \includegraphics[]{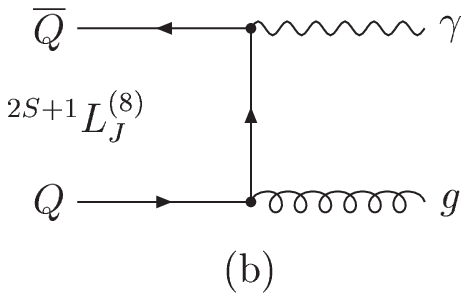}
 \end{center}
 \caption{Direct contributions to the radiative decay of a $Q\overline{Q}$
 	  pair: on the left (a) one of six colour singlet diagrams,
               on the right (b) one of two diagrams per colour octet mode.}
 \label{fig1}
\end{figure}

Taking only the leading order in $v/c$ the NRQCD result coincides with the one
of the CSM. However, subleading terms in the velocity expansion could be still
important numerically: In radiative decays the partonic kernel of a colour
octet contribution is enhanced by an inverse power of $\alpha_s(m_Q)$ with
respect to the leading order colour singlet mode. While the leading term needs
two hard gluons in the final state (cf fig.~\ref{fig1}(a)) a $Q\overline{Q}$
pair in a colour octet state can annihilate into a photon and a single gluon
(cf fig.~\ref{fig1}(b)). Thus one may also take into account the subleading
terms in $v/c$.

The photon energy spectrum in the hard subprocess $Q\overline{Q}[n] \to \gamma
X$ of radiative decays has been calculated in next-to-leading order
perturbative QCD for both the colour singlet mode \cite{Kramer:1999bf} and the
colour octet modes \cite{Maltoni:1999nh}. Another perturbative contribution may
become important in the upper endpoint region ($z = 2E_\gamma/M_H \to 1$) of
the spectrum. Due to an imperfect cancellation between terms stemming from real
and virtual emission of soft gluons one could expect potentially large
logarithms $\ln(1 - z)$ to all orders of the perturbation theory. A resummation
of these logarithms would then give rise to a Sudakov suppression $\sim
\exp(-\alpha_s \ln^2(1 - z) )$. Though an earlier analysis
\cite{Photiadis:1985hn} claimed such a Sudakov damping factor in the colour
singlet mode a more recent work \cite{Hautmann:1997sb} predicts such Sudakov
factors in the colour octet channels only while the logarithms should cancel
order by order $\alpha_s$ in the colour singlet mode.

\begin{figure}[t]
\begin{center}
 \leavevmode
 \includegraphics[]{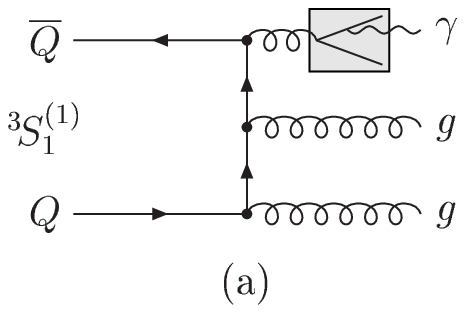}
 \qquad
 \includegraphics[]{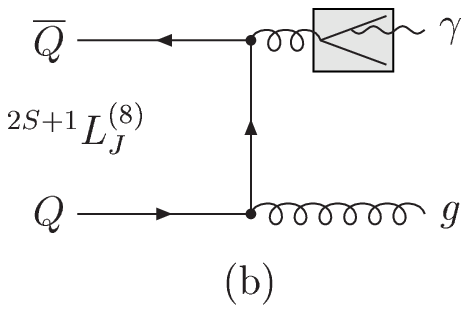}
 \qquad
 \includegraphics[]{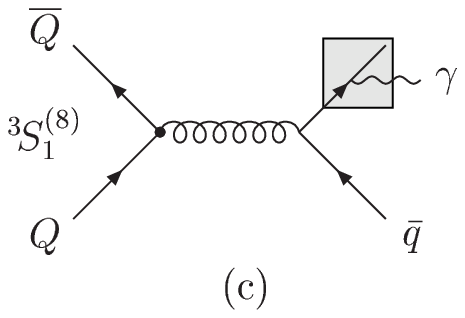}
 \end{center}
 \caption{Fragmentation contributions to the radiative decay of a
	  $Q\overline{Q}$ pair: colour singlet contribution (a), colour octet
	  contribution (b) and (c). The non-perturbative subprocesses inside
          the grey boxes are described by the gluon (a) + (b) and the quark (c)
          fragmentation function respectively.}
 \label{fig2}
\end{figure}

Besides these perturbative contributions several non-perturbative effects have
been investigated as well. They become important near the phase space
boundaries where the photon energy fraction $z$ is small or close to 1,
respectively. For low values of $z$ there is a large fragmentation contribution
caused by the collinear emission of a photon from a light (anti)quark in the
final state. Examples for such processes are diagrammed in fig.~\ref{fig2}.
They have been investigated in \cite{Catani:1995iz} for the colour singlet mode
and in \cite{Maltoni:1999nh} for the colour octet channels.

At the upper endpoint of the spectrum two different sources for
non-perturbative effects exist. The first one is the phase space effect
associated with the hadronization of massless gluons into massive final
states. This effect usually is considered by a parton shower Monte Carlo
thereby generating a non-zero invariant mass for the outgoing gluon(s)
\cite{Field:1983cy}. Another method based on the introduction of an effective
gluon mass \cite{Parisi:1980jy} could obtain the appropriate phase space
suppression by fitting the values of the effective gluon mass to data of
radiative $J/\psi$ \cite{Abrams:1980sbRonan:1980czScharre:1981yn} and
$\Upsilon$ \cite{Schamberger:1984mmCsorna:1986ivAlbrecht:1987hzBizzeti:1991ze,
Nemati:1997xy} decays independently \cite{Consoli:1994ewConsoli:1997ts}.

In this article we concentrate on another non-perturbative effect contributing
to the upper endpoint of the spectrum. In this region NRQCD operators connected
to the center-of-mass (cms) movement of the $Q\overline{Q}$ pair inside the
quarkonium could become significant even though they are subleading in the
sense of the naive NRQCD power counting \cite{Mannel:1997uk, Rothstein:1997ac}.
It has been shown in \cite{Rothstein:1997ac} explicitly that the NRQCD
velocity expansion breaks down near the endpoint. The reason for this breakdown
is the kinematical enhancement of the cms operators. In the endpoint region the
expansion parameter is $v^2/\epsilon$ rather than $v^2$ where $\epsilon = 1 -
E_\gamma/m_Q$ is a measure for the distance from the endpoint. Thus the
velocity expansion works fine only for photon energies that are significantly
further away from the endpoint than $\Delta E_\gamma \sim m_Q v^2$. However,
the range of applicability of NRQCD can be extended to higher values for
$E_\gamma$ by the resummation of an infinite class of operators into so-called
shape functions \cite{Rothstein:1997ac}. Thereafter the shape function improved
spectrum holds up to a resolution of $\Delta E_\gamma \sim m_Q v^2$.

The shape function formalism yields different shape functions for the different
quantum states $n$ of the quark-antiquark pair ($Q\overline{Q}[n]$). In the
leading colour singlet case the shape function is nothing else than a {\it
number} namely the corresponding NRQCD matrix element. In the colour octet
modes the shape functions are real {\it functions} reflecting the phase space
dependence of the soft gluon radiation. Thus the partonic spectrum of the
relevant channels $n = \{ {}^1\!S_0^{(8)}, {}^3\!P_J^{(8)}, {}^3\!S_1^{(8)} \}$
which is proportional to $\delta(E_\gamma - m_Q)$ is smeared out to a quite
broad peak.

As mentioned above their contribution may be as large as the one from the
leading order. If we compare the total rates of the leading colour singlet and
a subleading colour octet term we get
\begin{equation}
\label{rough}
\Gamma_{\!1} : \Gamma_{\!8} = \frac{\alpha_s(m_Q)}{4\pi} : v^4 = {\cal O}(1)\,.
\end{equation}
Thus it is worth to model the non-perturbative shape functions to estimate
their influence on the photon energy distribution. For the construction of the
model we keep close to a model successfully used for shape functions in
quarkonium production \cite{Beneke:2000gq}. In this case the shape functions
which originally were defined in \cite{Beneke:1997qw} have been modeled by soft
gluon emission from the final state. Inspired by the success in describing the
$J/\psi$ production in $B$ decays and in the photoproduction channel we take
over the physically simple picture of radiating off soft gluons from the heavy
(anti)quark.

We will proceed as follows: First we will recapitulate the result of underlying
partonic process $Q\overline{Q}[n] \to \gamma + g$ and $Q\overline{Q}[n] \to
\gamma + g + g$ for $n = {}^{2S+1}L_J^{(C)} \in \{ {}^1\!S_0^{(8)},
{}^3\!P_J^{(8)}, {}^3\!S_1^{(8)} \}$ and $n = {}^3\!S_1^{(1)}$
respectively. Moreover we will show to what extent fragmentation contributions
must be taken into account. Afterwards the construction and application of our
shape function model in the decay mode is given. Finally we discuss the results
of the numerical evaluation of the semi-inclusive decay $\Upsilon$(1S) $\to
\gamma$ + {\it light hadrons}.

\section{The partonic calculation}

Within NRQCD the photon energy spectrum in the semi-inclusive decay $H \to
\gamma X$ of a quarkonium is represented by the operator product expansion
\begin{equation}
\label{OPE}
\frac{d\Gamma}{d\hat{z}} = \sum_n {\cal C}[n] \bra{H} {\cal O}[n] \ket{H} \,.
\end{equation}
Here the Wilson coefficient ${\cal C}[n]$ is calculable perturbatively and
gives the differential rate $d\Gamma_{\!n}/d\hat{z}$ in the decay of a
$Q\overline{Q}$ pair with quantum numbers $n$ into a photon and light hadrons
$X$. Note that in conventional NRQCD the photon energy is normalized on the
quark rather than on the quarkonium mass: $\hat{z} = E_\gamma/m_Q$.

\subsection{Direct contributions}

Equation (\ref{OPE}) includes not only the contributions from the direct
production of a photon (fig.~\ref{fig1}(a)) but also the production via
fragmentation (fig.~\ref{fig1}(b)). We will come to this point later. We first
deal with the direct channels.  The leading term in the non-relativistic
expansion is the colour singlet mode displayed in diagram \ref{fig1}(a). It has
been calculated perturbatively up to ${\cal O}(\alpha_s)$ \cite{Kramer:1999bf}.
For the sake of simplicity we restrict ourselves on the tree level
contribution. It is \cite{Brodsky:1978du}
\begin{equation}
\label{Cdir1}
\begin{split}
{\cal C}_\gamma^{\rm dir}[^3\!S_1^{(1)}](\hat{z})
& = \frac{32 e_Q^2 \alpha_{\rm em}(m_Q) \alpha_s^2(m_Q)}{27 m_Q^2} \bigg[
        \frac{2-\hat{z}}{\hat{z}} + \frac{\hat{z}(1-\hat{z})}{(2-\hat{z})^2}
\\
& \hspace*{7em} + 2 \frac{1-\hat{z}}{\hat{z}^2} \ln(1-\hat{z})
                - 2 \frac{(1-\hat{z})^2}{(2-\hat{z})^3} \ln(1-\hat{z}) \bigg] .
\end{split}
\end{equation}

The subleading terms ${\cal O}(v^4)$ in the velocity expansion arise from
Feynman diagrams like fig.~\ref{fig1}(b), where ${}^{2S+1}L_J^{(C)} =
{}^1\!S_0^{(8)}$, ${}^3\!P_0^{(8)}$ or ${}^3\!P_2^{(8)}$. They are
perturbatively enhanced in comparison to the colour singlet channel (${\cal
O}(\alpha_s)$ versus ${\cal O}(\alpha_s^2)$). Due to their two body kinematics
the photon spectrum is fixed to a definite energy value:
\begin{equation}
\label{Cdir}
{\cal C}_\gamma^{\rm dir}[n](\hat{z})
= \frac{1}{2 \cdot 2m_Q} \, \frac{1}{8\pi} \,
H[Q\overline{Q}[n] \to \gamma g](2m_Q) \, \delta(1-\hat{z}) \,.
\end{equation}
The spin and colour averaged squares $H[Q\overline{Q}[n] \to \gamma g](2m_Q)$
of the amplitudes for the three relevant channels are (again we take only
leading terms in $\alpha_s$) \cite{Maltoni:1999nh}:
\begin{subequations}
\label{Hdir8}
\begin{align}
H[Q\overline{Q}[^1\!S_0^{(8)}] \to \gamma g](2m_Q)
& = \frac{256 \pi^2 e_Q^2 \alpha_{\rm em}(m_Q) \alpha_s(m_Q)}{2m_Q} \,,
\\
H[Q\overline{Q}[^3\!P_0^{(8)}] \to \gamma g](2m_Q)
& = \frac{768 \pi^2 e_Q^2 \alpha_{\rm em}(m_Q) \alpha_s(m_Q)}{2m_Q} \,,
\\
H[Q\overline{Q}[^3\!P_2^{(8)}] \to \gamma g](2m_Q)
& = \frac{1024 \pi^2 e_Q^2 \alpha_{\rm em}(m_Q) \alpha_s(m_Q)}{5 \cdot 2m_Q}\,.
\end{align}
\end{subequations}

\subsection{Fragmentation contributions}
\label{hardfrag}

Let us turn to the fragmentation contributions now. The corresponding processes
are associated with diagrams like the ones in fig.~\ref{fig2}. There the photon
does not stem directly from the annihilation of the heavy quark-antiquark pair
but from the fragmentation of a gluon or a light (anti)quark in the final
state. Nevertheless we can keep the form of equation (\ref{OPE}) to describe
these contributions, since this subprocess is independent from the initial
state effects parameterized in the NRQCD matrix elements. The Wilson
coefficient is then obtained by the folding
\begin{equation}
\label{Cfrag}
{\cal C}_\gamma^{\rm frag}[n](\hat{z}) =
\sum_{a = g, q, \bar{q}} \int\limits_{\hat{z}}^1 \!\! \frac{d\hat{x}}{\hat{x}}
        \, {\cal C}^{\rm dir}_a[n](\hat{x},\mu^2)
        \, D_{a\to\gamma}(\hat{z}/\hat{x},\mu^2)
\end{equation}
of the fragmentation function $D_{a\to\gamma}$ and the coefficient ${\cal
C}^{\rm dir}_a[n]$ which is the perturbative part of the NRQCD decay rate of a
$Q\overline{Q}$ pair with quantum numbers $n$ into a particle $a \in \{ g, q,
\bar{q} \}$ and other light degrees of freedom. The integration variable
$\hat{x}$ indicates the energy of the particle $a$ normalized on the heavy
quark mass: $\hat{x} = E_a/m_Q$.

Since the fragmentation takes place at a scale far below the heavy quark mass
$m_Q$ it can be factorized from the hard subprocess. This is denoted by the
factorization scale $\mu$ in (\ref{Cfrag}). As usual one may derive a
renormalization group equation from the $\mu$ independence of ${\cal
C}_\gamma^{\rm frag}[n]$ to shift potentially large logarithms between ${\cal
C}^{\rm dir}_a[n]$ and $D_{a\to\gamma}$.

Though the fragmentation functions are non-perturbative objects a naive
estimate for their order of magnitude is obtained from counting coupling
constants and collinear singularities coming up in a perturbative
calculation. For that we look at the subprocesses highlighted by grey boxes in
fig.~\ref{fig2}. The coupling of the photon to a light (anti)quark is
proportional to $\alpha_{em}(m_Q)$. For $D_{q\to\gamma}$ the leading term comes
from the kinematic region where the photon and the (anti)quark are
collinear. Thus one gets a factor $\ln(Q^2/Q_0^2)$ where $Q^2 \sim m_Q^2$ from
the phase space integration with a collinear cut-off parameter $Q_0$ of order
$\Lambda_{\rm QCD}$. Accordingly one gets $\alpha_{em}(m_Q) \alpha_s(m_Q)
\ln^2(Q^2/Q_0^2)$ for the gluon fragmentation function. Here the logarithm
appears quadratically because quark, antiquark, and photon can become collinear
simultaneously in this case.

The logarithms $\ln(Q^2/Q_0^2)$ could become so large that they could
compensate the perturbative $\alpha_s$ suppression and thus confuse the
perturbation series dramatically. This is seen in a easy way from the running
of the strong coupling constant. At leading order the renormalization group
equation yields
\begin{equation}
\alpha_s(\mu^2) = \frac{\alpha_s(\mu_0^2)}
    {1 + \alpha_s(\mu_0^2) \frac{\beta_0}{4\pi} \ln(\frac{\mu^2}{\mu_0^2})}
\end{equation}
with $\beta_0 = (33 - 2n_f)/3$ if $n_f$ fermions are active. As long as $\mu$
is far above a typical hadronic scale $\mu_0 \sim \Lambda_{\rm QCD}$ one can
neglect the constant in the denominator. Hence
\begin{equation}
\label{LLA}
\alpha_s(\mu^2) \ln(\mu^2/\mu_0^2) \sim {\cal O}(1) \,.
\end{equation}
Based on this relation we receive fragmentation contributions with magnitudes
comparable to the direct ones. This become clear if we rewrite the
fragmentation functions in the {\it leading logarithmic approximation}. Then
they have the form \cite{Owens:1987mp}
\begin{equation}
D_{a\to\gamma}(\xi,Q^2)
= \frac{2}{\beta_0} \frac{\alpha_{\rm em}(Q^2)}{\alpha_s(Q^2)} f_a(\xi)
\end{equation}
where $f_a$ is a phenomenological function of the fraction $\xi = E_\gamma/E_a$
of the photon and the parton energy. In case of comparable functions $f_a$ for
the different partons $a$ the $\alpha_s$ in the denominator cancels the
additional $\alpha_s$ in the hard subprocess of the fragmentation
contributions.

Instead of using fitted functions for $f_a$ we will take the perturbative
result for the fragmentation functions $D_{a\to\gamma}$. They are known in
next-to-leading order \cite{Aurenche:1993ycGluck:1993zxBourhis:1998yu} but we
will restrict ourselves on the leading order for consistency with the
investigation of the annihilation subprocess. The scale dependence of
$D_{g\to\gamma}$ and $D_{q\to\gamma}$ is given by the leading order DGLAP
equations \cite{Gribov:1972riGribov:1972rtLipatov:1975qmDokshitzer:1977sg,
Altarelli:1977zs}
\begin{subequations}
\begin{align}
\label{DGLAP}
\mu^2 \frac{\partial}{\partial\mu^2} D_{q \to \gamma}(\xi, \mu^2)
& = \frac{e_q^2 \alpha_{\rm em}(\mu^2)}{2 \pi} P_{q \to \gamma}(\xi) \,,
\\
\mu^2 \frac{\partial}{\partial\mu^2} D_{g \to \gamma}(\xi, \mu^2) & =
\frac{\alpha_s(\mu^2)}{2 \pi} \int\limits_\xi^1 \! \frac{d\eta}{\eta} \,
                P_{g \to q}(\eta) \, D_{q \to \gamma}(\xi/\eta,\mu^2)
\end{align}
\end{subequations}
where $e_q$ is the charge of the light quark measured in units of the
elementary charge. The Altarelli--Parisi splitting functions are
\cite{Altarelli:1977zs}:
\begin{equation}
P_{q \to \gamma}(\xi) = \frac{1 + (1-\xi)^2}{\xi} \,,
\qquad
P_{g \to q}(\xi) = \frac{1}{2} [\xi^2 + (1-\xi)^2] \,.
\end{equation}

Integrating equation (\ref{DGLAP}) one obtains the quark fragmentation function
\begin{equation}
\label{Dqtogam0}
D_{q \to \gamma}(\xi, \mu^2)
= \frac{e_q^2 \alpha_{\rm em}(\mu^2)}{2 \pi} P_{q \to \gamma}(\xi)
  \, \ln\!\left(\frac{\mu^2}{\mu^2_0 (1-\xi)^2}\right)
+ D_{q \to \gamma}(\xi, \mu_0^2) \,.
\end{equation}
The $\xi$ dependence of the starting value $D_{q \to \gamma}(\xi, \mu_0^2)$ for
the evolution in the factorization scale $\mu$ is mainly determined by a
logarithm $\ln(1/(1 - \xi)^2)$ originating from the phase space integration
\cite{Glover:1994xc}. It is already separated in (\ref{Dqtogam0}). The
remaining rest term cannot be calculated perturbatively. Therefore it has to
be modeled, e.g.~with a vector dominance model, or fitted to data. The latter
was done by the ALEPH collaboration in a measurement of the $\gamma$ + (1 jet)
rates for $\xi > 0.7$ \cite{Buskulic:1996au}. They get the best fit for a
constant rest term $C = - 1 - \ln(M_Z^2/(2\mu_0^2))$ in
\begin{equation}
\label{Dqtogam}
D_{q \to \gamma}(\xi, \mu^2)
= \frac{e_q^2 \alpha_{\rm em}(\mu^2)}{2 \pi} \left[
    P_{q \to \gamma}(\xi) \, \ln\!\left(\frac{\mu^2}{\mu_0 (1-\xi)^2}\right)
  + C \right] .
\end{equation}
The non-perturbative scale $\mu_0$ extracted from data is then
\begin{equation}
\mu_0 = 0.14^{+0.43}_{-0.12} \, \text{GeV} \quad \Rightarrow \quad
C = - 13.26^{+2.81}_{-3.89} \,.
\end{equation}

The non-perturbative piece of the gluon fragmentation function $D_{g \to
\gamma}$ cannot be determined experimentally. For the sake of simplicity we
will set $D_{g \to \gamma}(\xi, \mu_0) = 0$. The leading quadratic logarithmic
term reads
\begin{equation}
\label{Dgtogam}
D_{g \to \gamma}(\xi, \mu^2)
= \frac{\alpha_s(\mu^2)}{2 \pi} \int\limits_\xi^1 \! \frac{d\eta}{\eta} \,
        P_{g \to q}(\eta) \, D_{q \to \gamma}(\xi/\eta, \mu^2) \,
        \frac{1}{2} \ln\!\left(\frac{\mu^2}{\mu^2_0}\right) .
\end{equation}

Since the experimental investigation of the photon spectrum in radiative
quarkonia decays is limited to $z > 0.4$ due to large uncertainties for soft
photons caused by $\pi^0$ decays we need the fragmentation functions for $\xi >
0.4$ only. In this region the contribution of the gluon fragmentation function
is negligible. However, the quark fragmentation into a photon is significant
for $\xi \sim 0.4$. Furthermore possibly large contributions in the endpoint
region we are especially interested in are caused by the logarithmic divergence
$\ln(1/(1 - \xi)^2)$.

At this stage one comment is in order. Since our approximation for the
fragmentation functions and in particular relation (\ref{LLA}) holds the better
the larger the scale $\mu$ is, i.e.~the larger the heavy quark mass $m_Q$ is,
the leading log approximation is inaccurate or even not reliable for $m_Q =
m_c$. Therefore we will concentrate our numerical investigation on the
$\Upsilon$ decay as long as we do not restrict ourselves on large values for
$z$.

Finally we need the perturbative results for the coefficient ${\cal C}^{\rm
dir}_a[n]$ in (\ref{Cfrag}). Again we distinguish between colour singlet and
colour octet contributions. The calculation of the colour singlet mode
(fig.~\ref{fig2}(a) without the fragmentation subprocess) is, except for colour
factors, the same as the decay rate of ortho-positronium \cite{Ore:1949te}.
Similarly the coefficient ${\cal C}^{\rm dir}_a[^3\!S_1^{(1)}]$ can be
extracted from the photon energy spectrum (\ref{Cdir}). The relative factor
\begin{equation}
B_F = \frac{\sum_{abc} (\frac{T_F}{2} d^{abc})(\frac{T_F}{2} d^{abc})}
        {\sum_{ab} (\frac{1}{2} \delta^{ab})(\frac{1}{2} \delta^{ab})}
= \frac{N_c^2 - 4}{4 N_c} \stackrel{(N_c=3)}{=} \frac{5}{12}
\end{equation}
gives the ratio of the corresponding colour traces. Inclusive of the coupling
constants the triple gluon energy spectrum in the decay
$Q\overline{Q}[^3\!S_1^{(1)}] \to g g g$ results in
\begin{equation}
{\cal C}^{\rm dir}_g[^3\!S_1^{(1)}](\hat{x})
= B_F \frac{\alpha_s(m_Q)}{e_Q^2 \alpha_{\rm em}(m_Q)} \,
        {\cal C}_\gamma^{\rm dir}[^3\!S_1^{(1)}](\hat{x}) \,.
\end{equation}
This formula contains a combinatorial factor $1/3 = 1/3! : 1/2!$ which
compensates the aforementioned factor of three coming from the fact that all
three final state gluons can fragment into a photon.

The colour octet coefficients ${\cal C}_a^{\rm dir}[^{2S+1}L_J^{(8)}]$ are
determined by the hard subprocesses in diagrams like fig.~\ref{fig2}(b) and
fig.~\ref{fig2}(c). In both cases the kinematics are trivial. Thus we get for
${\cal C}^{\rm dir}_g[n]$ with $n \in \{ {}^1\!S_0^{(8)}, {}^3\!P_0^{(8)},
{}^3\!P_2^{(8)} \}$ from diagram~\ref{fig2}(b):
\begin{equation}
{\cal C}^{\rm dir}_g[n](\hat{x})
= 2 \, \frac{1}{2 \cdot 2m_Q} \, \frac{1}{8\pi}
        \, H[Q\overline{Q}[n] \to g g](2m_Q) \, \delta(1 - \hat{x}) \,.
\end{equation}
Up to a factor of two which again is caused by the possibility for both gluons
to fragment into a photon ${\cal C}^{\rm dir}_g[n](\hat{x})$ matches the
corresponding differential decay modes $d\hat{\Gamma}_{\!n}/d\hat{x}$ without
the NRQCD matrix element. The spin and colour averaged square of the amplitudes
are \cite{Maltoni:1999nh}:
\begin{subequations}
\begin{align}
H[Q\overline{Q}[^1\!S_0^{(8)}] \to g g](2m_Q)
& = B_F \, \frac{128 \pi^2 \alpha_s^2(m_Q)}{2m_Q} \,,
\\
H[Q\overline{Q}[^3\!P_0^{(8)}] \to g g](2m_Q)
& = B_F \, \frac{384 \pi^2 \alpha_s^2(m_Q)}{2m_Q} \,,
\\
H[Q\overline{Q}[^3\!P_2^{(8)}] \to g g](2m_Q)
& = B_F \, \frac{512 \pi^2 \alpha_s^2(m_Q)}{5 \cdot 2m_Q} \,.
\end{align}
\end{subequations}

Final states with $J = 1$ are forbidden by the Landau-Yang theorem
\cite{Landau:1948Yang:1950rg}, i.e.~$n = {}^3\!P_1^{(8)}$ and $n =
{}^3\!S_1^{(8)}$ do not contribute to $Q\overline{Q}[n] \to g g$. Instead the
latter configuration can decay into a light quark-antiquark pair
(cf fig.~\ref{fig2}(c)). The corresponding coefficient is
\begin{equation}
{\cal C}^{\rm dir}_a[^3\!S_1^{(8)}](\hat{x})
= \frac{1}{2 \cdot 2m_Q} \, \frac{1}{8\pi} \,
     H[Q\overline{Q}[^3\!S_1^{(8)}] \to q\bar{q}](2m_Q) \, \delta(1 - \hat{x})
\end{equation}
where $a \in \{ q, \bar{q} \}$ and
\begin{equation}
H[Q\overline{Q}[^3\!S_1^{(8)}] \to q\bar{q}](2m_Q)
= \frac{64 \pi^2 \alpha_s^2(m_Q)}{3 \cdot 2m_Q} \,.
\end{equation}

Hence we have collected all the results for the partonic decay of a
$Q\overline{Q}$ pair that are needed for the photon energy distribution in
radiative decays of heavy quarkonia. Henceforth we will construct a model for
shape functions in quarkonia decays and embed the partonic results into it.

\section{The shape function model}

\begin{figure}[t]
  \begin{center}
    \leavevmode
    \includegraphics[]{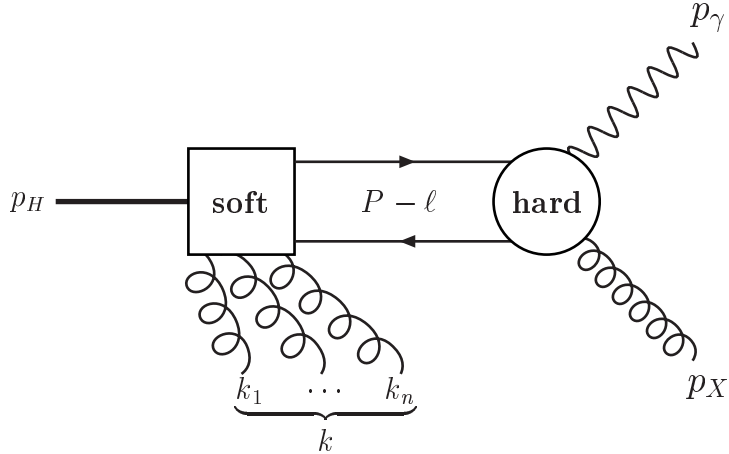}
  \end{center}
  \vspace{-3ex}
  \caption{Schematic representation of direct colour octet contributions to the
           radiative quarkonium decay.}
  \label{fig3}
\end{figure}

In the following both direct and fragmentation contributions are improved by
the application of the shape functions. We model these functions with the
following physical picture in mind. According to the factorization assumption
of NRQCD quarkonium decays are divided into two subprocesses as illustrated in
fig.~\ref{fig3}. In the first stage the quarkonium $H$ radiates off a cluster
of gluons with a momentum $k = \sum_i k_i \sim {\cal O}(m_Q v^2)$. The final
state of this non-perturbative subprocess is a $Q\overline{Q}$ pair in the
quantum state $n$ that annihilates perturbatively in the second stage. Its
momentum is given by $p_{Q\overline{Q}} = P - \ell$ with $P^2 = (2m_Q)^2$,
i.e.~the non-perturbative momentum $\ell \sim {\cal O}(\Lambda_{\rm QCD}$)
measures the off-shellness of the quark-antiquark pair.

It is important to be careful with neglecting non-perturbative momenta in the
hard subprocess because the light cone component $\ell_+$ of the
$Q\overline{Q}$ off-shellness is responsible for the shift of the partonic
endpoint $m_Q$ to the physically correct hadronic value $E_\gamma^{\rm max} =
M_H/2$ \cite{Rothstein:1997ac}. Therefore we will keep the kinematics exactly
here.

Nevertheless we have to model the radiation of the gluons from the initial
state. This is done by our shape function \cite{Beneke:2000gq}
\begin{equation}
f^H_n(\ell) = \int \! \frac{d k^2}{2 \pi} \frac{d^3 \bs{k}}{(2\pi)^3 2 k^0} \,
              (2\pi)^4 \delta^4(p_H + k - P + \ell) \, \Phi_n(k; p_H, P)
\end{equation}
where $\Phi_n$ is a radiator function parameterizing the emission of a soft
gluon cluster with total momentum $k$. In our ansatz
\begin{equation}
\label{ansatz}
\Phi_n(k;p_H,P) = a_n \cdot |\bs{k}|^{b_n} \exp\{ - k_0^2 / \Lambda_n^2 \}
                      \cdot k^2 \exp\{ - k^2 / \Lambda_n^2 \}
\end{equation}
the cut-off parameter $\Lambda_n \sim {\cal O}(m_Q v^2)$ reflect the
expectation that the main contribution comes from the ultrasoft region where
the energy $k_0$ and the invariant mass $k^2$ of the gluon cluster are of order
$m_Q v^2$ and $(m_Q v^2)^2$ respectively. The choice
\begin{subequations}
\label{paras}
\begin{align}
& b[{}^1\!S_0^{(8)}] = 2 \,, \qquad
        b[{}^3\!P_0^{(8)}] = b[{}^3\!S_1^{(8)}] = 0 \,,
\\
& \Lambda[{}^1\!S_0^{(8)}] = \Lambda[{}^3\!P_0^{(8)}] \equiv \Lambda \,,
\quad \Lambda[{}^3\!S_1^{(8)}] = c \Lambda
\end{align}
\end{subequations}
for the constants in (\ref{ansatz}) is motivated by the fact that the gluon
coupling for a M1 magnetic dipole transition from the quarkonium $H$ to
$Q\overline{Q}[^1\!S_0^{(8)}]$ is proportional to the gluon momentum while an
E1 or a double E1 electric dipole transition to
$Q\overline{Q}[^3\!P_{0,2}^{(8)}]$ or $Q\overline{Q}[^3\!S_1^{(8)}]$
respectively does not have any $k$ dependence. Furthermore the necessity of at
least two transitions for $n = {}^3\!S_1^{(8)}$ suggests the introduction of a
factor $c = 1.5$ to enlarge the average radiated energy and invariant mass in
this case. Finally we fix $a_n$ by the normalization condition
\begin{equation}
\label{shapenorm}
\int \! \frac{d^4\ell}{(2\pi)^4} \, f^H_n(\ell)
= \frac{1}{(2\pi)^3} \int\limits_0^\infty \!\! dk^2 \!\!
                \int\limits_{\sqrt{k^2}}^\infty \!\! dk_0 \,
                \sqrt{k_0^2-k^2} \, \Phi_n(k; p_H, P)
= \bra{H} {\cal O}_n \ket{H} \,.
\end{equation}

\subsection{Direct contributions}

After we have determined our model ansatz for the shape functions we proceed
with the implementation of the hard subprocess. First we deal with the direct
contributions. As mentioned above in leading order only the colour octet modes
are interesting in the shape function formalism. Thus we start with the
expression
\begin{equation}
\begin{align}
d\Gamma_{\!8}^{\rm dir}
& = \sum_n \int \! \frac{d^4\ell}{(2\pi)^4}
        \frac{1}{2M_H} \int \!\! \widetilde{dp}_\gamma \, \widetilde{dp}_X
        (2\pi)^4 \delta^4(P - \ell - p_\gamma - p_X) \,
        H[Q\overline{Q}[n] \!\to\! \gamma g](P, \ell, p_\gamma, p_X)
\nonumber \\
&\hspace{8em}
\cdot \int \! \frac{d k^2}{2 \pi} \frac{d^3\bs{k}}{(2\pi)^3 2 k_0} \,
        (2\pi)^4 \delta^4(p_H - k - P + \ell) \, \Phi_n(k; p_H ,P)
\label{dGam8dir}
\end{align}
\end{equation}
where one can easily recognize the shape function in the second line of the
equation. In (\ref{dGam8dir}) $M_H$ denotes the quarkonium mass,
$\widetilde{dp}_\gamma$ and $\widetilde{dp}_X$ the invariant phase space
measures of the photon and the hard gluon respectively and finally the spin and
colour averaged square $H_n$ of the hard subprocess amplitude is given by
(\ref{Hdir8}).

Manipulating eq.~(\ref{dGam8dir}) we start with integration out the four
momenta $k$ and $p_X$ with $p_X^2 = 0$ which determines the light cone
components $\ell_\perp^2$ and $\ell_+$ by the delta functions to
\begin{subequations}
\label{ellcone}
\begin{align}
\ell_\perp^2
& = (M_H - 2m_Q)(M_H - 2m_Q + 2\ell_0) + \ell_+(2\ell_0 - \ell_+) - k^2 \,,
\\
\ell_+
& = \frac{1}{2E_\gamma} \left[
  4m_Q (M_H - E_\gamma) - M_H^2 - 2(M_H - 2E_\gamma) \ell_0 + k^2 \right] .
\end{align}
\end{subequations}
To make use of these relations we also decompose $d^4\ell$ into its light cone
components
\begin{equation}
\label{dell}
\int \!\! d^4\ell = \int\limits_0^{2\pi} \!\! d\phi \,
\int \!\! d\ell_0 d\ell_+ \frac{d\ell_\perp^2}{2} \, \Theta(\ell_\perp^2)
\end{equation}
and rewrite the $\ell_0$ integration into a $k_0$ integration by $\ell_0 = k_0
- (M_H - 2m_Q)$. Then the integration over the azimuthal angular can be
performed trivially because the partonic process is $\phi$ independent.
Analogously the angular dependence in $\widetilde{dp}_\gamma$ is integrated out
trivially and one gets $\widetilde{dp}_\gamma = E_\gamma dE_\gamma
\Theta(E_\gamma)/(4 \pi^2)$. Finally we evaluate the equations (\ref{ellcone}).
In simultaneous consideration of the theta function in (\ref{dell}) there arise
integration bounds for $k_0$ where the physical ones are
\begin{equation}
\frac{(M_H - 2E_\gamma)^2 + k^2}{2 (M_H - 2E_\gamma)}
\le k_0 \le \frac{M_H^2 + k^2}{2 M_H} .
\end{equation}
From this we can read off an upper bound for the $k^2$ integration
\begin{equation}
k^2 \le M_H (M_H - 2E_\gamma)
\end{equation}
while the lower one is determined by $k^2 \ge 0$. Introducing the abbreviations
\begin{equation}
\label{albet}
\alpha = (p_H - p_\gamma)_+ = M_H - 2E_\gamma \,, \qquad
\beta  = (p_H - p_\gamma)_- = M_H
\end{equation}
we finally end up with
\begin{equation}
\label{dGdEdir8}
\frac{d\Gamma_{\!8}^{\rm dir}}{dE_\gamma}
= \sum_n \int\limits_0^{\alpha\beta} \frac{dk^2}{2\pi} \!
\int\limits_{(\alpha^2+k^2)/(2\alpha)}^{(\beta^2+k^2)/(2\beta)} \hspace{-2em}
        dk_0 \,\, \frac{1}{2M_H} \!\cdot\! \frac{1}{8\pi} \,
        H[Q\overline{Q}[n] \to \gamma g](M_{Q\overline{Q}}(k))
        \, \cdot \, \frac{1}{4\pi^2} \, \Phi_n(k; p_H) \,.
\end{equation}
This is our master equation for the direct colour octet contributions to the
photon energy spectrum that take into account shape functions effects within
our model framework. Note that according to our model the partonic rate
depends on
\begin{equation}
M_{Q\overline{Q}}(k) = \sqrt{M_H^2 - 2 M_H k_0 + k^2}
\end{equation}
rather than on $2m_Q$, i.e.~the quark mass in the partonic subprocess is
effectively larger than $m_Q$. The colour singlet contributions are obtained by
multiplying (\ref{Cdir1}) with the colour singlet NRQCD matrix element $\bra{H}
{\cal O}_1({}^3\!S_1) \ket{H}$. Here the heavy quark mass is set equal to
$M_H/2$.

\subsection{Fragmentation contribution}

The treatment of the fragmentation contributions is done in the following way:
First we apply our shape function model on the colour octet contribution
$d\hat{\Gamma}/d\hat{E}_a(Q\overline{Q}[n] \to a \orbar{a})$ to extend the
reliability of the partonic NRQCD calculation up to higher values of the parton
energy $\hat{E}_a$ in the $Q\overline{Q}$ rest frame. Afterwards we fold the
received spectrum $d\Gamma/dE_a$ with the corresponding fragmentation function
$D_{a \to \gamma}$:
\begin{equation}
\label{dGamdzfrag}
\frac{d\Gamma_{\!8}^{\rm frag}}{dE_\gamma}
= \sum_{a = g, q, \bar{q}} \int\limits_{E_\gamma}^{M_H/2} \!\! \frac{dE_a}{E_a}
  \, \frac{d\Gamma_{\!8}^{\rm dir}}{dE_a} \, D_{a \to \gamma}(E_\gamma/E_a) \,.
\end{equation}
Note that the upper bound for the parton energy $E_a$ in the quarkonium rest
frame is given by the hadronic value $M_H/2$ rather than by the heavy quark
mass $\hat{E}^{\rm max}_a = m_Q$ which defines the endpoint in the partonic
calculation.

In (\ref{dGamdzfrag}) the folding of the gluon and the (anti)quark energy
spectrum $d\Gamma_{\!8}^{\rm dir}/dE_a$ is calculated completely analogous to 
(\ref{dGdEdir8}):
\begin{displaymath}
\frac{d\Gamma_{\!8}^{\rm dir}}{dE_a}
= \sum_n \int\limits_0^{\bar{\alpha}\bar{\beta}} \frac{dk^2}{2\pi} \!
\int\limits^{(\bar{\beta}^2+k^2)/(2\bar{\beta})}
           _{(\bar{\alpha}^2+k^2)/(2\bar{\alpha})} \hspace{-2em}
        dk_0 \,\, \frac{1}{2M_H} \!\cdot\! \frac{1}{8\pi} \,
        H[Q\overline{Q}[n] \to a \orbar{a}](M_{Q\overline{Q}}(k))
        \, \cdot \, \frac{1}{4\pi^2} \, \Phi_n(k; p_H)
\end{displaymath}
where the sum again runs only over the colour octet modes. $\bar{\alpha}$ and
$\bar{\beta}$ are obtained from (\ref{albet}) by the substitution $E_\gamma \to
E_a$. As described in subsection \ref{hardfrag} the partonic subprocess
$Q\overline{Q}[n] \to g g$ contributes only for $n = {}^1\!S_0^{(8)}$, $n =
{}^3\!P_0^{(8)}$, and $n = {}^3\!P_2^{(8)}$ while $n = {}^3\!S_1^{(8)}$ needs a
light quark-antiquark pair in the final state. The sum in (\ref{dGamdzfrag})
runs not only over the gluon but also over different (anti)quark flavours $q =
\{ u, d, s (, c) \}$ where we additionally assume $D_{q \to \gamma}(\xi) =
D_{\bar{q} \to\gamma}(\xi)$.

\section{Results}

As mentioned above we restrict our numerical analysis on the bottomonium system
since the fragmentation contributions in the charmonium sector are not very
trustworthy within our approach. For the photon spectrum in the radiative decay
of a $\Upsilon$(1S) we choose the following set of parameters: $M_\Upsilon =
9.46$ GeV, $\alpha_s(\mu^2) = 0.190$, $\alpha_{\rm em}(\mu^2) = 1/132$ where
the factorization scale $\mu$ is fixed at the $b$ quark mass $m_b = 4.8$
GeV. Unfortunately the values of the NRQCD matrix elements are unknown. While
the production matrix elements of the bottomonium sector have been fitted in a
recent analysis by Braaten, Fleming and Leibovich \cite{Braaten:2000cm} we have
to make do with the NRQCD scaling rules
\begin{equation}
\label{vregel}
\bra{\Upsilon} {\cal O}_8(^1\!S_0) \ket{\Upsilon}
\sim \frac{\bra{\Upsilon} {\cal O}_8(^3\!P_0) \ket{\Upsilon}}{m_b^2}
\sim \bra{\Upsilon} {\cal O}_8(^3\!S_1) \ket{\Upsilon}
\sim v^4 \bra{\Upsilon} {\cal O}_1(^3\!S_1) \ket{\Upsilon}
\end{equation}
to estimate at least the order of magnitude of the decay matrix elements needed
for the normalization, i.e.~for the relative weights, of the different
contributing channels. For our numerical evaluation we take for each colour
octet matrix element $2.20 \cdot 10^{-2}$ $\text{GeV}^2$ and $\bra{\Upsilon}
{\cal O}_1(^3\!S_1) \ket{\Upsilon} = 3.43$ $\text{GeV}^2$, i.e.~we set $v^2 =
0.08$ for bottomonia. Finally we normalize the total rate to one. In case of a
comparison with data this could be changed easily.

\begin{figure}[t]
 \begin{center}
  \includegraphics[width=0.45\textwidth, bb=105 220 470 495]{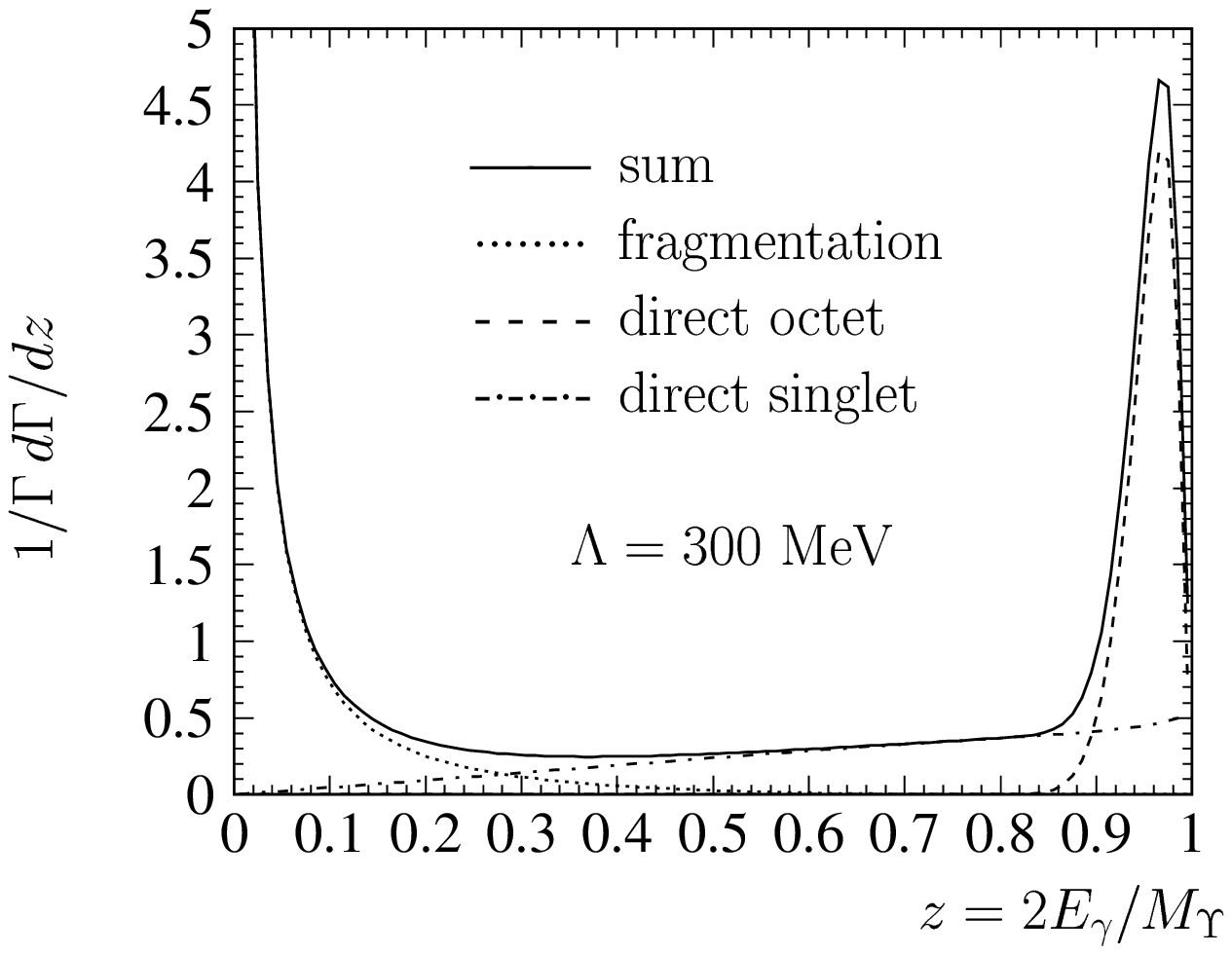}
  \quad
  \includegraphics[width=0.45\textwidth, bb=105 220 470 495]{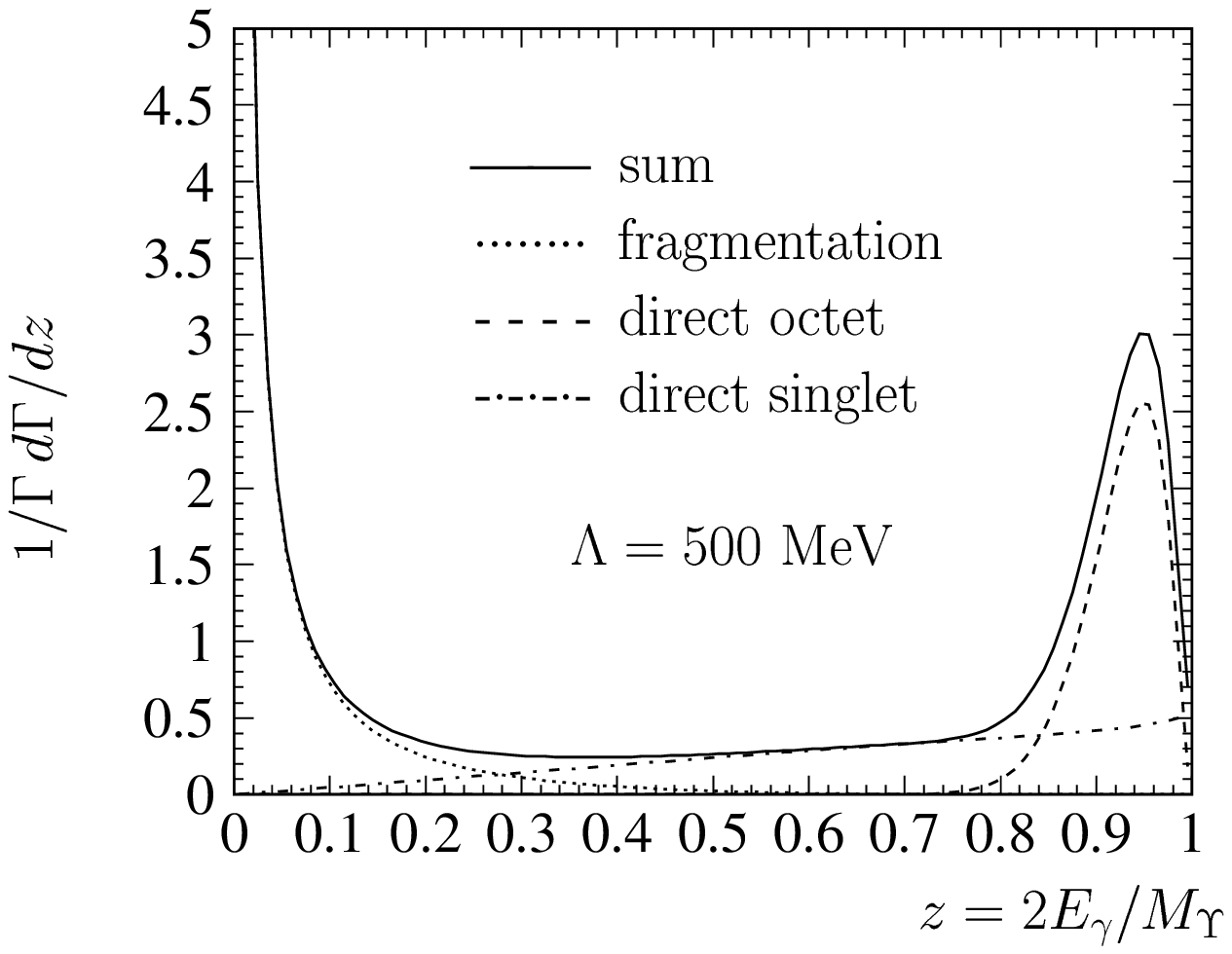}
 \end{center}
 \vspace{-9ex}
 \caption{Photon energy spectrum in the radiative decay $\Upsilon(\text{1S})
  \to \gamma$ + {\it light hadrons} for two different values of the shape
  function model parameter: $\Lambda = 300$ MeV (left) and $\Lambda = 500$ MeV
  (right).}
  \label{fig4}
\end{figure}

The result for two different values of our shape function model parameter
$\Lambda \sim m_b v^2$ is show in fig.~\ref{fig4}. One recognizes that
fragmentation contributions are non-negligible for small photon energies
only. In the region $0.1 \lesssim z \lesssim 0.4$ they are mainly dominated by
the $^3\!S_1^{(8)}$ channel. Though this channel is connected with the
fragmentation function $D_{q \to \gamma}(\xi)$ which diverges for $\xi \to 1$
its numerical contribution to the spectrum for large values of $z$ is
negligible. This matches to our expectation that fragmentation processes prefer
to transfer small energy fractions from the gluon (quark) to the photon.

The upper endpoint region $z \gtrsim 0.8$ is dominated by direct colour octet
contributions (at least in leading order $\alpha_s$). While the partonic result
is proportional to $\delta(1 - \hat{z})$ the soft gluon radiation in the
initial state smears out the delta peak and also shifts its maximum to
$z_{Q\overline{Q}} = M_{Q\overline{Q}}^{\rm eff}/M_\Upsilon < 1$. The higher
the model parameter $\Lambda$ is the smaller is $z_{Q\overline{Q}}$ and
therewith the effective heavy quark mass and the broader is the width of the
peak. Comparing the integrated rates of the direct colour singlet mode
$^3\!S_1^{(1)}$ and the direct colour octet ones we realize that their
contributions are almost equal. This can be explained by the fact that the
suppression factor $v^4 = 6.4 \cdot 10^{-3}$ of the colour octet modes is
canceled by an additional factor $\alpha_s(m_b)/(4\pi) = 1.5 \cdot 10^{-2}$ in
the colour singlet mode (cf equation (\ref{rough})). Furthermore one have to
consider that there are contributions from many colour octet modes
(${}^{2S+1}L_J^{(C)} = {}^1\!S_0^{(8)}, {}^3\!S_1^{(8)}, {}^3\!P_0^{(8)}$ and
${}^3\!P_2^{(8)}$) but only from one colour singlet mode (${}^3\!S_1^{(1)}$).

Let us finally concentrate on the region of middle high photon energies. From
the theoretical point of view the part between $0.4 \lesssim z \lesssim 0.75$
is the cleanest one of the spectrum. Here the colour singlet contribution which
is assumed to be dominating over the whole photon energy range in the colour
singlet model can be extracted within shape function improved NRQCD without any
pollution from other channels. This is still true after including the
next-to-leading order contributions: While the colour singlet corrections
steepen the slope \cite{Kramer:1999bf} the colour octet terms are negligible
for $0.4 \lesssim z \lesssim 0.75$ \cite{Maltoni:1999nh}.

\begin{figure}[t]
 \begin{center}
  \includegraphics[width=0.45\textwidth, bb=105 220 470 495]{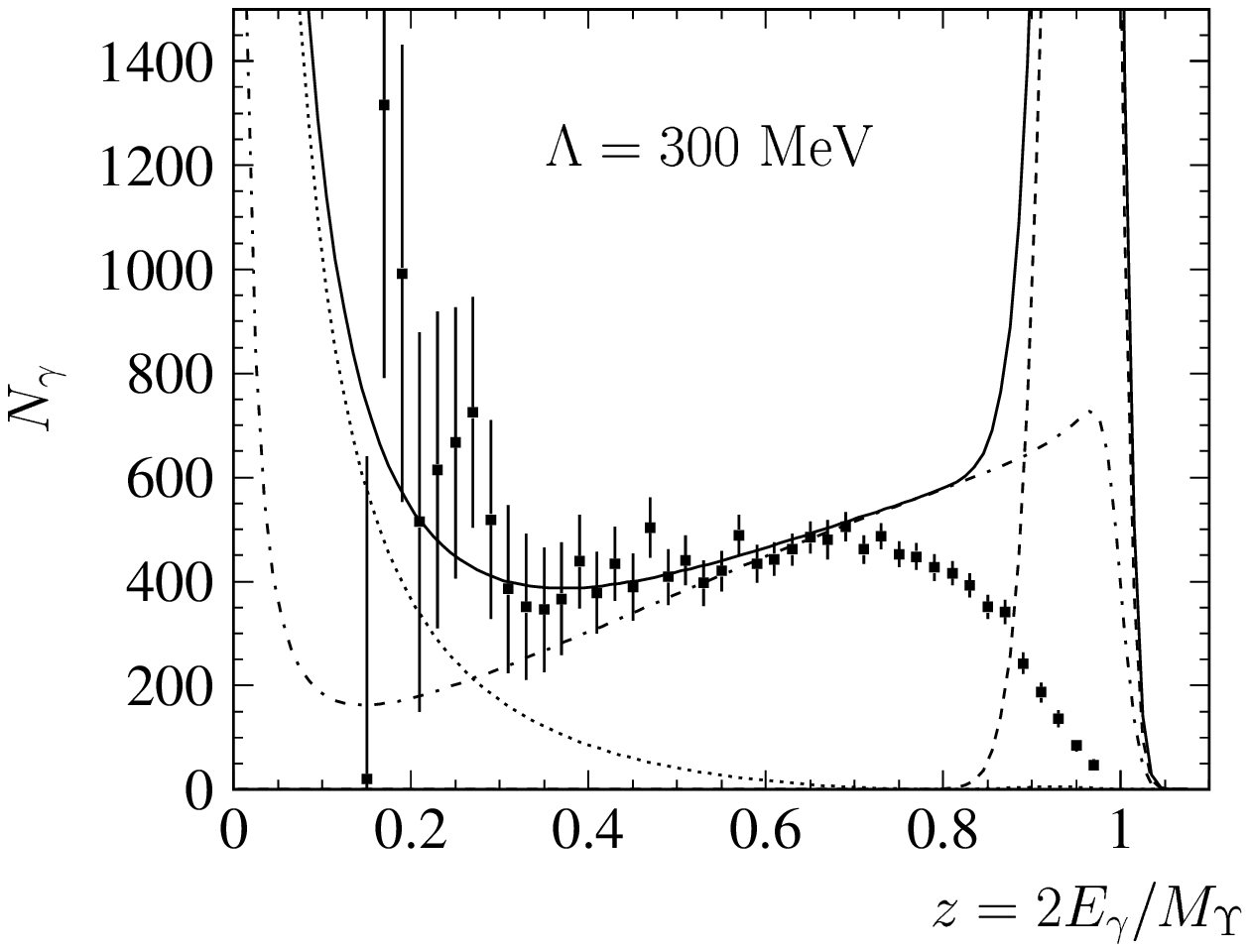}
  \quad
  \includegraphics[width=0.45\textwidth, bb=105 220 470 495]{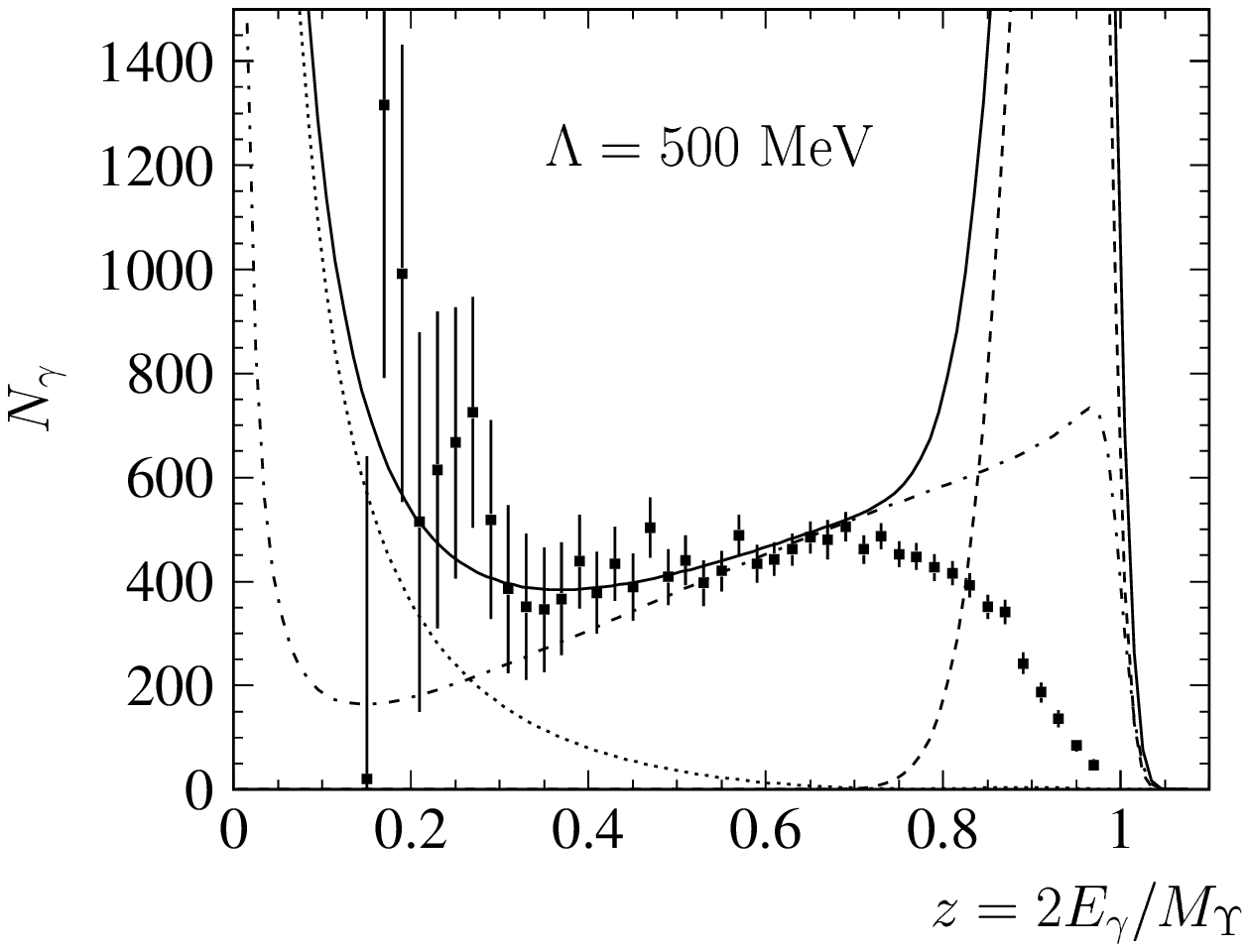}
 \end{center}
 \vspace{-2ex}
 \caption{Comparison of the theoretical spectrum (solid line) for $\Lambda =
  300$ MeV (left) and $\Lambda = 500$ MeV (right) with data of CLEO
  \cite{Nemati:1997xy}. The dashed line shows the direct, the dotted line the
  fragmentation contribution. The spectrum predicted by the colour singlet
  model (direct + fragmentation, without any hadronization model) is indicated
  by the dash-dotted curvature.}
  \label{fig5}
\end{figure}

Since the extraction of $\alpha_s(m_b)$ needs an extrapolation of the measured
spectrum towards $z = 0$, a complete theoretical understanding of the part of
the spectrum the fit is based on is indispensable to reach an accurate value
for the strong coupling constant. In our opinion this is not possible for high
photon energies. To illustrate the problems in the upper endpoint region we
smear out our theoretical result with the energy resolution of the CLEO
detector
\begin{equation}
\frac{\sigma_E}{E}(\%) = \frac{0.35}{E^{0.75}} + 1.9 - 0.1E
\end{equation}
and fit it to their most recent data \cite{Nemati:1997xy} for $0.4 \le z \le
0.7$. The result is shown in fig.~\ref{fig5}. While the spectrum is described
satisfactorily within the fragmentation uncertainties for small and middle high
values of $z$ the discrepancy between theory and experiment is overwhelming for
$z \gtrsim 0.75$. As shown by the CLEO collaboration \cite{Nemati:1997xy} the
CSM result combined with the fragmentation contributions according to
\cite{Catani:1995iz} can be brought into agreement with data using a
hadronization model by Field \cite{Field:1983cy} even though the consequential
value of $\alpha_s(m_b)$ is slightly to small compared to the one measured on
the $Z_0$ resonance.

Although the colour octet contributions in fig.~\ref{fig5} have not been
suppressed by a hadroni\-zation model yet, they seem to be in strong
contradiction to the experimental observation. The simplest explanation for
this deviation would be an extreme smallness of the colour octet NRQCD decay
matrix elements even smaller than the $v^4$ suppression still acknowledged by
the power counting rules. Some hints for such small colour octet decay matrix
elements also come from the estimation of the $\alpha_s$ corrections
\cite{Maltoni:1999nh}. Furthermore a recent analysis of the corresponding
production matrix elements \cite{Braaten:2000cm} yielded smaller values than
predicted by velocity scaling rules, too. Although the crossing symmetry
between these production matrix elements and the ones of the decay holds only
in leading order perturbation theory this could also be interpreted as
indication for somehow suppressed (or even negative?) values of the NRQCD
matrix elements.

Nevertheless the understanding of the upper endpoint region in the radiative
decay of the $\Upsilon$(1S) is not good enough to conclude convincingly that
the colour octet matrix elements are extremely small. Without having
investigated the Sudakov corrections on the colour octet contributions and
without a better understanding of the hadronization process it seems impossible
to give a stringent theoretical prediction for $z \gtrsim 0.75$. Furthermore
the experimental investigation of this spectrum suffers from large systematic
problems in this kinematical regime, too.

In summary an extraction of a precise value for the strong coupling constant
from the radiative $\Upsilon$ decay seems to be impossible as long the upper
endpoint region is included in the fit. Unless both theoretical and
experimental progress concerning the physics in the upper endpoint region is
achieved $\alpha_s(m_b)$ should be fitted from the data between $0.4 \le z \le
0.7$ only.

\subsection*{Acknowledgements}
I would like to thank M.~Beneke for useful comments and for reading the
manuscript. The author is supported by the Graduiertenkolleg
``Elementarteilchenphysik an Beschleunigern'' and the DFG-Forschergruppe
``Quantenfeldtheorie, Computeralgebra und Monte-Carlo-Simulation''.

\end{document}